# Manifestation of the collective drift of molecules in argon according to their mean square displacements

Nikolay P. Malomuzh[1)], Konstantin S. Shakun[2, a)]

[1)]Odessa National University, Dept. of Theor. Phys., 2, Dvorjanskaya str., Odessa, 65026, Ukraine

[2)]Odessa National Maritime Academy, Dept. of Phys., 8, Didrikhson str., Odessa, 65029, Ukraine

The mean square displacement (MSD) of an argon molecule as a function of time is studied. Its deviations from the standard asymptotic law for intermediate times are analyzed in details. It is shown that these deviations are mainly connected with the square-root contribution to the MSD which is proportional the ratio of the collective part to the full self-diffusion coefficient. It is established that the relative value of the collective contribution to the self-diffusion coefficient of argon changes from 0.23 near the triple point up to 0.4 at approaching the critical point. A new method for the determination of the Maxwell relaxation time is proposed. Its temperature dependence on the coexistence curve and one of isochors is investigated.

---

[a] Author to whom correspondence should be addressed. Electronic mail: *gluon2008@meta.ua*

# I. INTRODUCTION

Liquid occupies the intermediate position between solids and gases. At that, all molecular motions in solids have the collective nature and, vice versa, all motions in rare gases are individual. Therefore it is natural to expect that the thermal motion of molecules in liquids should be some combination of collective and individual constituents. This conclusion is supported in particular by the long-living tells of the velocity autocorrelation function (VACF) Refs. 1, 2. In accordance with Refs. 1, 2 the behavior of the VACF for long times is described by the power law:

$$\phi_{\vec{V}}(t) = <\vec{V}(t)\vec{V}(0)> \underset{t\to\infty}{\to} \frac{A}{t^{3/2}}. \tag{1}$$

From the physical point of view, such an asymptote of the VACF is caused by the long-living transversal hydrodynamic modes. There are several independent theoretical approaches explaining their manifestation in the behavior of the VACF Refs. 3-12.

The approach proposed by I. Fisher Ref. 12 occupies the special position. Within it, the motion of a molecule is considered as the superposition of collective and individual motions. The peculiar attention is focused on the collective motion which is associated with the hydrodynamic motion of a Lagrange particle. In accordance with that the self-diffusion coefficient $D_s$ of a molecule appears to be the sum

$$D_s = D_c + D_r. \tag{2}$$

of collective $D_c$ and individual $D_r$ terms. The last contribution is genetically connected with those displacements usually considered in the kinetic theory of gases and liquids Ref. 11. The first contribution is less standard. It is tightly connected with the fluctuation vortexes which generate the long time tails (1) of the VACF of a molecule Refs. 12, 13, 14-16, 18. Sizes of these vortexes are essentially bigger in comparison with the molecular size. A vortex 1 displaces a molecule on small distance 1, after some time a vortex 2, arising near vortex 1, displaces the same molecule on new small segment 2 and so on (see Fig.1). This new mechanism of transport in liquids is only characteristic for liquids. It can be weakly manifested in dense gases too. It is absent in solids and rare gases. The contribution $D_c$ has different character of the temperature dependence in comparison with $D_r$ and can be separated Refs. 18-20.

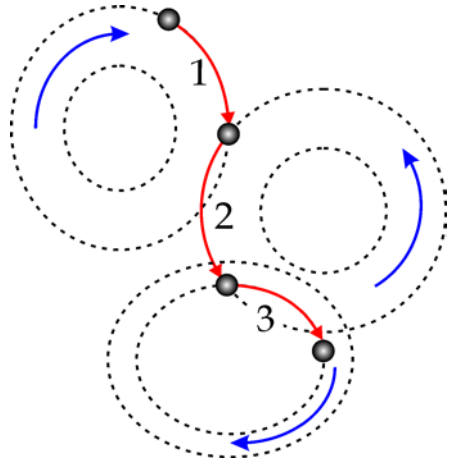

Fig. 1. Schematic picture for the collective transport created by fluctuation vortexes in liquid (vortex motion is set by arrows).

The consecutive study of the collective transport, as well as of the ratio $D_c/D_s$, had been the object of careful study in Refs. 13-21. It had been shown that $D_c/D_s$ changes from $0.05 \div 0.2$ near the triple points of liquids up to $0.4 \div 0.5$ in the vicinity of the critical point. However, these estimates essentially depend on the values of the Maxwell relaxation time. The last had been studied only for several liquids and several temperatures Refs. 22-25. Therefore, the further study of this question seems to us very important.

The present paper is devoted to the careful study of time dependence for the mean square displacement $\Gamma(t)$ of a molecule in liquid argon. We want to investigate the contribution to $\Gamma(t)$ caused by hydrodynamic modes, responsible for the asymptote (1). We will show that the analysis of this contribution to $\Gamma(t)$ allows us to get an independent estimate for the ratio $D_c/D_s$. The consistency of this approach with direct calculation of $D_c/D_s$ is discussed. A new method for the determination of the Maxwell relaxation time will be proposed.

## 2. MANIFESTATION OF THE COLLECTIVE DRIFT IN THE BEHAVIOR OF $\Gamma(t)$

In accordance with Ref. 12 the velocity of a molecule is represented as a sum of two terms:

$$\vec{V}(t) = \vec{V}_L(t) + \vec{V}_r(t). \tag{3}$$

The first of them is the velocity of a Lagrange particle, including the present molecule and its surroundings, and the second is the relative velocity of the same particle to its nearest neighbors. With suitable accuracy both these velocities propose to be statistically independent. Taking into account this circumstance and for the VACF of a molecule $\phi_{\vec{V}}(t)$ we can write:

$$\phi_{\vec{V}}(t) = \phi_L(t) + \phi_r(t), \tag{4}$$

where $\phi_L(t) = <\vec{V}_L(t)\vec{V}_L(0)>$ and $\phi_r(t) = <\vec{V}_r(t)\vec{V}_r(0)>$. Supposing that the drift of a Lagrange particle in the field of thermal hydrodynamic fluctuations is an essentially slower process than the displacement of a molecule relative to its nearest neighbors, one can conclude that

$$\phi_{\vec{V}}(t) \underset{t\to\infty}{\to} \phi_L(t). \tag{5}$$

The VACF for a Lagrange particle is calculated with the help of the Lagrange theory of thermal hydrodynamic fluctuations proposed in Ref. 12 and developed in Refs. 13-18. Here we take into account, that the mean square displacement $\Gamma_L(t)$ of a Lagrange particle and its VACF are connected by the equation

$$\Gamma_L(t) = 2\int_0^t du(t-u)\phi_L(u) \quad \text{or} \quad \frac{d^2\Gamma_L(t)}{dt^2} = 2\phi_L(t). \tag{6}$$

Using the results of the Lagrange theory of thermal hydrodynamic fluctuation and equations (1) and (5) for large enough times we can find:

$$\Gamma_L(t) = 6D_L t - \frac{k_B T}{\rho[\pi(\nu+D_L)]^{3/2}}\sqrt{t} + \ldots, \tag{7}$$

where $D_L = \frac{1}{3}\int_0^\infty \phi_L(t)dt$ is the self-diffusion coefficients of a Lagrange particle. The square root term in (6) is generated by the long time asymptote (1).

The equation for $\Gamma_r(t)$ is similar to (5), so it leads to the result:

$$\Gamma_r(t) = 6D_r t + C,$$

where $D_r = \frac{1}{3}\int_0^\infty \phi_r(t)dt$ is the contribution to the self-diffusion coefficient caused by the relative motion of a molecule, $C = -2\int_0^\infty t\phi_r(t)dt$. The function $\phi_r(t)$ is different from zero only for small and intermediate times and the main contribution into $C$ is connected with that part of the function under integral where it is negative, so we expect that $C$ is positive.

In accordance with the above, the mean square displacement of a molecule is represented in form:

$$\Gamma(t) = \Gamma_L(t) + \Gamma_r(t) \Rightarrow C + 6D_s t - \frac{k_B T}{\rho[\pi(\nu+D_L)]^{3/2}}\sqrt{t} + \ldots, \tag{8}$$

where $D_s = D_c + D_r$ is the full self-diffusion coefficient of a molecule. This expression for $\Gamma(t)$ is consistent with $\phi_L(t)$ obtained first in Refs. 12-14:

$$\phi_L(t) = \frac{2k_B T}{\pi\rho}\frac{1}{(4(\nu+D_L)t)^{3/2}} \Rightarrow \frac{A}{t^{3/2}}, \quad A = \frac{2k_B T}{m}\tau_0^{3/2},$$

where $\tau_0 = \dfrac{1}{(8\pi n)^{2/3} v} = \dfrac{1}{(8\pi)^{2/3}} \dfrac{a^2}{v}$ and $a$ is the interparticle spacing. The inequality

$\dfrac{D_L}{v} \leq \dfrac{D_c}{v} < \dfrac{D_s}{v} \ll 1$ is also used. For argon near the triple point $\tau_0 \sim 10^{-13}$ s.

It is necessary to note that the account of the square root contribution to $\Gamma(t)$ is only correct for

$$t > (\gg) t_0, \quad t_0 = \dfrac{r_p^2}{\pi v} \approx \tau_0.$$

The inequality $\dfrac{k_B T}{\rho(\pi v)^{3/2}} \sqrt{t} < \dfrac{1}{n} 6 D_s t$, where $n$ is an integer, takes place for $t > n^2 t_0$. It means, that the simplified asymptote $\Gamma_{simp}(t) = C + 6 D_s t$ will describe the MSD with the accuracy $0.99$ only for $t > 10^4 t_0 \approx 10^3$ ps. This circumstance is very important for the correct determination of $D_s$ and $C$.

### a) Collective contribution to the self-diffusion coefficient

The collective part of the self-diffusion coefficient is connected with one for the Lagrange particle with the relation:

$$D_c = D_L \big|_{r_L = r_*}, \tag{9}$$

where $r_*$ is "the suitable radius" of a Lagrange particle. According to Ref. 18,

$$r_* = 2\sqrt{v \tau_M}, \tag{10}$$

where $v$ is the kinematic shear viscosity and $\tau_M$ is the Maxwell relaxation time (MRT). In connection with this we note that the MRT is defined as the characteristic time for the frequency dependent shear viscosity of liquids:

$$v(\omega) = \dfrac{v}{1 + i\omega \tau_M}, \quad v = v(\omega = 0).$$

From here it follows that

$$\nu(\omega) = \begin{cases} \nu, & \omega\tau_M \ll 1, \\ G_\infty/i\omega\rho, & \omega\tau_M \gg 1, \end{cases}$$

where $G_\infty = \nu\rho/\tau_M$ is the high-frequency shear elasticity modulus Ref. 22. Thus, the MRT divides the reaction of liquids on viscous one for low frequencies and elastic one for high frequencies. Such a character of the shear viscosity leads to the following hydrodynamic equation for the transversal hydrodynamic velocity $\vec{u}$ (see below and Ref. 23:

$$\frac{\partial \vec{u}}{\partial t} + \tau_M \frac{\partial^2 \vec{u}}{\partial t^2} = \nu\Delta\vec{u}, \qquad (11)$$

that describes the viscous relaxation of liquid for slow processes and its elastic evolution for rapid ones.

As a result $D_c$ appears to be equal (see Ref. 18):

$$D_c = \frac{k_B T}{10\pi\eta\sqrt{\nu\tau_M}}. \qquad (12)$$

Combining (8) and (11) we get:

$$\Gamma(t) = C + 6D_s t \left[ 1 - \frac{10}{3\pi^{1/2}} \frac{D_c}{D_s} \left(\frac{\tau_M}{t}\right)^{1/2} + \ldots \right]. \qquad (13)$$

Introducing the dimensionless variables: $x = t/\tau_M$, $\tilde{\Gamma}(x) = \Gamma(t)/6D_s\tau_M$ and $\tilde{C} = C/6D_s\tau_M$ we can rewrite (13) in the form:

$$\frac{D_c}{D_s} = \frac{3\pi^{1/2}}{10} x^{1/2} \left[ 1 - \frac{\tilde{\Gamma}(x) - \tilde{C}}{x} \right] + o(1/x^{1/2}).$$

Since the inequality $x^{1/2} \gg 1$ is consistent with $\tilde{\Gamma}(x) \gg \tilde{C}$, the last equation can be simplified:

$$\frac{D_c}{D_s} = F(x) + o(1/x^{1/2}), \quad F(x) = \frac{3\pi^{1/2}}{10} x^{1/2} \left[ 1 - \frac{\tilde{\Gamma}(x)}{x} \right]. \qquad (14)$$

In connection with this the ratio $\frac{D_c}{D_s}$ can be estimated as

$$\frac{D_c}{D_s} = AverF_{MD}(x),\tag{15}$$

where $F_{MD}(x) = \frac{3\pi^{1/2}}{10} x^{1/2}\left[1 - \frac{\tilde{\Gamma}_{MD}(x)}{x}\right]$ and $\tilde{\Gamma}_{MD}(x)$ is the MSD found in computer experiments and the symbol "*Aver*" denotes the averaging operation on computer simulation data errors in the behavior of $\tilde{\Gamma}_{MD}(x)$. It is supposed that $x$ is inside the time interval: $1 << x << x_c$, where $x_c$ is the upper limit for the applicability of $\tilde{\Gamma}_{MD}(x)$.

This limit is caused by the accumulation of computer integration errors. Let $\Delta(t) = <\left[(x_k^2(t) + y_k^2(t) + z_k^2(t)) - \tilde{\Gamma}_{MD}(t)\right]^2>_k$ be the mean square deviation of the square displacement for $k$-th trajectory from the MSD (the angular brackets denote the averaging on all computer simulation trajectories). Then the value $t_c$ is naturally determined by the equation: $\Delta(t_c) = e\Delta(1)$. Our analysis shows that $t_c > 100\,ps$ for all investigated temperatures.

Using the Einstein formula for the self-diffusion coefficient $D_s = \frac{k_B T}{6\pi\eta r_p}$, we can write:

$$\frac{D_c}{D_s} = \frac{3}{5}\frac{r_p}{\sqrt{\nu\tau_M}},\tag{16}$$

where $r_p$ is the effective radius of a molecule.

To find the ratio $\frac{D_c}{D_s}$ according to (15) it is necessary to find 1) the time dependence of the MSD and 2) the Maxwell relaxation time.

## b) New methods for the determination of the Maxwell relaxation time

In accordance with Refs. 3-16 the velocity of auto-correlation function $\varphi_{\vec{V}}(t)$ of a molecule for long times takes the asymptote:

$$\varphi_{\vec{V}}(t) \to \psi_{\vec{u}}(t), \tag{17}$$

where $\psi_{\vec{u}}(t)$ is the correlation function for transversal constituents of the hydrodynamic velocity field. It is connected with the spatial Fourier components $\vec{u}(\vec{k},t)$ by the formula:

$$\psi_{\vec{u}}(t) = \frac{1}{(2\pi)^3} \int_0^\infty <\vec{u}^*(\vec{k},t)\vec{u}(\vec{k},0)> 4\pi k^2 dk, \tag{18}$$

From the linearized Stokes equation Ref. 24 it follows that the Fourier component of transversal hydrodynamic velocity satisfies the equation:

$$\frac{\partial \vec{u}}{\partial t} = -\nu \vec{k}^2 \vec{u} \tag{19}$$

which leads to the solution:

$$\vec{u}(\vec{k},t) = \vec{u}(\vec{k},0)\exp(-\nu \vec{k}^2 t). \tag{20}$$

Since, $<\vec{u}^*(\vec{k},0)\vec{u}(\vec{k},0)> = \frac{2k_B T}{\rho}$ Ref. 25, from (18) and (20) we obtain:

$$\psi_{\vec{u}}^{(0)}(t) = \frac{2k_B T}{(4\pi\nu t)^{3/2}\rho}. \tag{21}$$

In more generalized case we can use the equation instead of (19):

$$\frac{\partial \vec{u}}{\partial t} + \tau_M \frac{\partial^2 \vec{u}}{\partial t^2} = -\nu \vec{k}^2 \vec{u}, \tag{22}$$

which corresponds to (11). The solution of (22) equals to

$$\vec{u}(\vec{k},t) = \vec{u}(\vec{k},0)\exp\left(-\frac{t}{2\tau_M}\left(1 - \sqrt{1-4\nu\tau_M \vec{k}^2}\right)\right).$$

From here and (18) we find:

$$\psi_{\vec{u}}(t) = \frac{k_B T}{\pi^2 \rho} \int_0^\infty \exp\left(-\frac{t}{2\tau_M}\left(1-\sqrt{1-4\nu\tau_M \vec{k}^2}\right)\right) k^2 dk. \tag{23}$$

For $t \gg \tau_M$ the main contribution in (23) is caused by wave vectors from the interval:

$\frac{1}{\nu t} \ll \vec{k}^2 \ll \frac{1}{\nu \tau_M}$. In this case the integrand can be essentially simplified:

$$\begin{aligned}\psi_{\vec{u}}(t) &= \frac{k_B T}{\pi^2 \rho} \int_0^\infty \exp(-\nu \vec{k}^2 t)\left[1 - \frac{t}{\tau_M}(\nu \tau_M \vec{k}^2)^2 + ...\right] k^2 dk \\ &\Rightarrow \left[1 - t\tau_M \frac{\partial^2}{\partial t^2} + ...\right]\psi_{\vec{u}}^{(0)}(t)\end{aligned} \tag{24}$$

The direct calculation leads to the final result:

$$\psi_{\vec{u}}(t) \Rightarrow \frac{A}{t^{3/2}}\left[1 - \frac{15}{4}\frac{\tau_M}{t} + ...\right], \qquad A = \frac{2k_B T}{(4\pi\nu)^{3/2}\rho}.$$

This expression allows us to propose a new method for the determination of the Maxwell relaxation time $\tau_M$:

$$\tau_M = \frac{4}{15}\lim_{t\to\infty} t\left(1 - \frac{\psi_{\vec{u}}(t) t^{3/2}}{A}\right), \tag{25}$$

or

$$\tau_M = \frac{4}{9}\lim_{t\to\infty} t\left(1 - \frac{\Gamma''_{MD}(t) t^{3/2}}{2A}\right). \tag{26}$$

It is evident, that our approach to the determination of the Maxwell relaxation time 1) is independent from that proposed in Refs. 26-28 and 2) allows us to verify its consistency with other branches of molecular physics.

## 3. DETERMINATION OF $\tau_M$ AND $D_c/D_s$ FOR ARGON

The computer simulations of the VACF for argon molecules for different temperatures are presented in the Appendix. Values of $\tau_M$ calculated in accordance with the formulas (25) and (26) are presented in the second line of the Table I. Values of the Maxwell relaxation time taken from other sources are given in other lines.

TABLE I. Values of the Maxwell relaxation time for liquid argon on its liquid-vapor coexistence curve

| $T, K$ | 90 | 100 | 110 | 120 | 130 | 140 | 150 |
|---|---|---|---|---|---|---|---|
| $\tau_M^{(VACF)} \cdot 10^{13}, s$ | 1.74 ± 0.16 | 1.67 ± 0.14 | 1.64 ± 0.09 | 1.62 ± 0.11 | 1.653 ± 0.09 | 1.771 ± 0.08 | 1.905 ± 0.06 |
| $\tau_M \cdot 10^{13}, s$ [22] | 1.68 | | 1.58 | 1.57 | 1.66 | 1.73 | |
| $\tau_M \cdot 10^{13}, s$ [23] | ≈2.28 | – | – | – | – | | |
| $\tau \cdot 10^{12}, s$ [24] | – | | ≈2.1 | ≈2.2 | – | – | |

As we see, there is quite satisfactory agreement of values for the MRT obtained in Ref. 26 and by us, although different methods are used. In Ref. 26 the MRT is recovered from the analysis of the autocorrelation function for shear stresses. However, the ensemble of 4000 particles seems to us too small for reliable description of the stress tensor correlations.

In the prior publications Refs. 27, 28 the standard Maxwell relation $\tau_M = \eta / G_\infty$ is applied and the main attention is focused on the determination of the high frequency shear modulus $G_\infty$. The last was connected in Ref. 27 with the increment of the potential energy of a system for high frequency shear strains. For the study of this question the computer simulations are used. The heuristic equality between $G_\infty$ and the bulk elasticity modulus $p_{th}/(T\gamma_p(T))$ in the high density range is used in Ref. 28 ( $p_{th} = T \left.\frac{\partial p}{\partial T}\right|_V$ is the thermal pressure and $\gamma_p(T)$ is the thermal expansion coefficient). Using experimentally determined values of $p_{th}$ the estimates for the MRT of several liquids had been obtained in Ref. 28.

It is necessary to note that our approach has one very important advantage before other methods: the analysis of the time dependence for the VACF of a molecule allows us to recover the self-diffusion coefficient, shear viscosity and Maxwell relaxation time. Therefore, we can control the accuracy of calculations for each of them.

Our results for the MRT on the coexistence curve are completed by values of the MRT on isochore: $\rho = 0.837 g/cm^3$.

TABLE II. Values of the MRT for liquid argon on its isochore $\rho = 0.837 g/cm^3$

| $T, K$ | 113 | 133 | 138 | 140 | 147 |
|---|---|---|---|---|---|
| $\tau \cdot 10^{13}, s$ | - | - | 2.26 ±0.09 | 2.11 ±0.07 | 2.14 ±0.11 |

Values of the ratio $\dfrac{D_c}{D_s}$ are determined as the averages of $F_{MD}(x)$ within time intervals where their lower and upper edges satisfy the inequality: $1 < x_l < x < x_u < x_c$. The values $x_l$ and $x_u$ should correspond to maximal deviations of $\Gamma_{MD}(x)$ from corresponding rectilinear asymptotes (see below). The behavior of $F_{MD}(x)$ inside such frames is presented in Fig. 2. For $x > x_c$ the relative role of square-root contribution become smaller and it leads to the decrease of $F_{MD}(x)$.

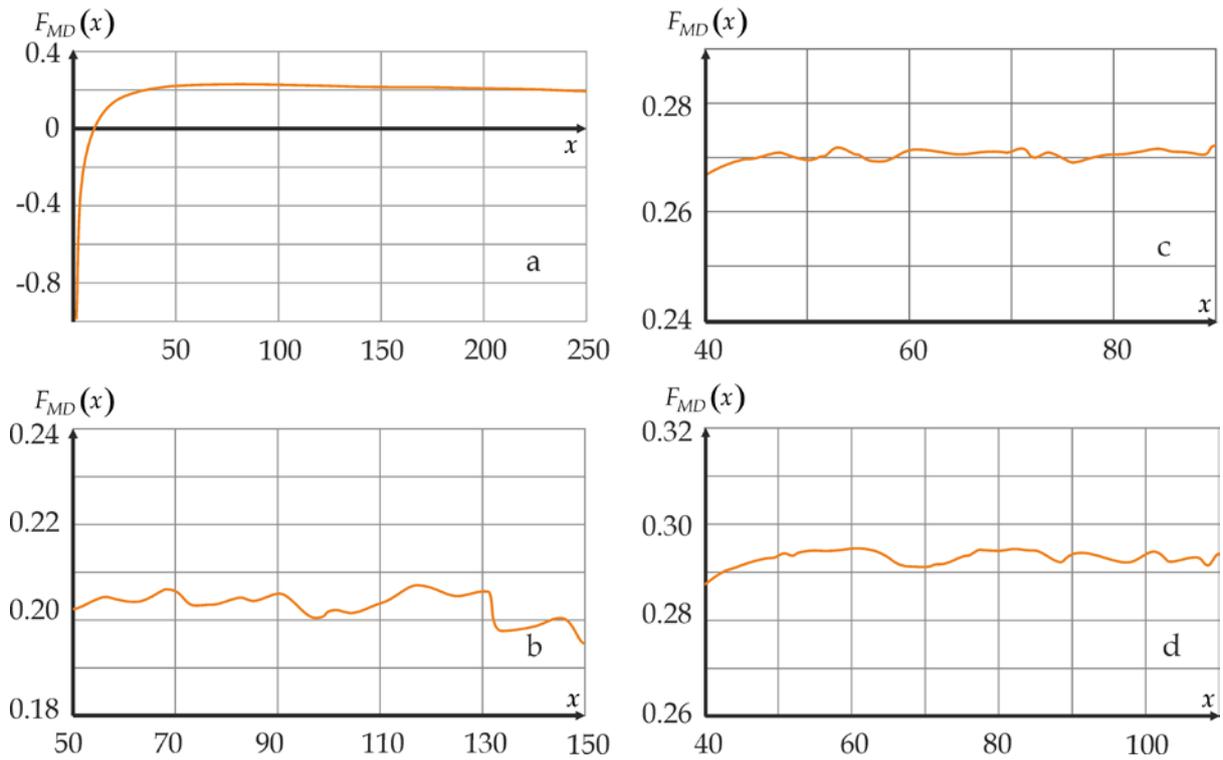

Fig.2. Behavior of smoothed $F_{MD}(x)$ for large enough $x$ at $T = 120K$ (a) and its behavior within time interval $x \in [x_l, x_u]$ for different temperatures:
$T = 100, 120, 130 K$ (b, c, d correspondingly).

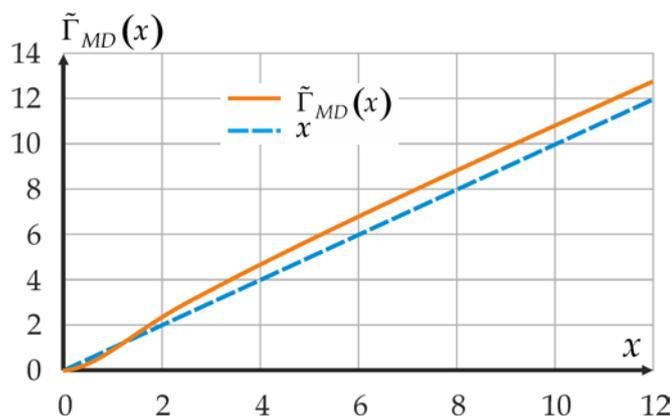

Fig. 3. The behavior of $\tilde{\Gamma}_{MD}(x)$ for $T = 90\ K$
(blue line corresponds to the equation $\tilde{\Gamma}(x) = x$).

Unfortunately, for $T < T_* \approx 93\ K$ the situation radically changes: for the large enough $x$ the difference $\tilde{\Gamma}_{MD}(x) - x$ becomes negative (see Fig. 3) while it is positive for all $T > T_*$. This fact has the simple qualitative explanation. For solid states, crystal or amorphous, the MSD is restricted quantity: $\tilde{\Gamma}_{MD}(x) < \Gamma_a$, where $\Gamma_a \sim \sigma^2$. It is clear that such a tendency should appear in liquid states at approaching the boundary between liquid and solid states. This question requires separate consideration that is planned to be made in another work. Here we only add that the more complicated behavior of $\tilde{\Gamma}_{MD}(x)$ is caused by all other terms that are not taken into account in the expansion (13) for $x \sim 1$.

The values of the ratio $\dfrac{D_c}{D_s}$ obtained in such a way are collected in the Table II. Values of $\dfrac{D_c}{D_s}$ calculated according to (15) and (10) are placed in the second column.

TABLE III. Comparative values of the ratio $D_c/D_s$ calculated according to (15) and immediately: $D_c$ are calculated according to (12) and $D_s$ are taken from the Table. AIII.

| $T, K$ | $D_c/D_s$, (14) | $D_c/D_s$ (11) |
|---|---|---|
| 40 | 0.042[a] | |
| 50 | 0.066[a] | |
| 60 | 0.125[a] | |
| 83.815 | 0.182 | |
| 90 | 0.189 | 0.223 |
| 100 | 0.204 | 0.231 |
| 110 | 0.234 | 0.262 |
| 120 | 0.27 | 0.302 |
| 130 | 0.293 | 0.363 |

[a]Values refer to the supercooled region where the procedure for determination of density is not quite reliable (see. Appendix)

Here we also see the systematic excess of the values of $D_c/D_s$ calculated directly, i.e. with the help of (11) and experimental values of the self-diffusion coefficient, over the values of $D_c/D_s$ obtained according to (15). The difference of these values does not exceed 20%.

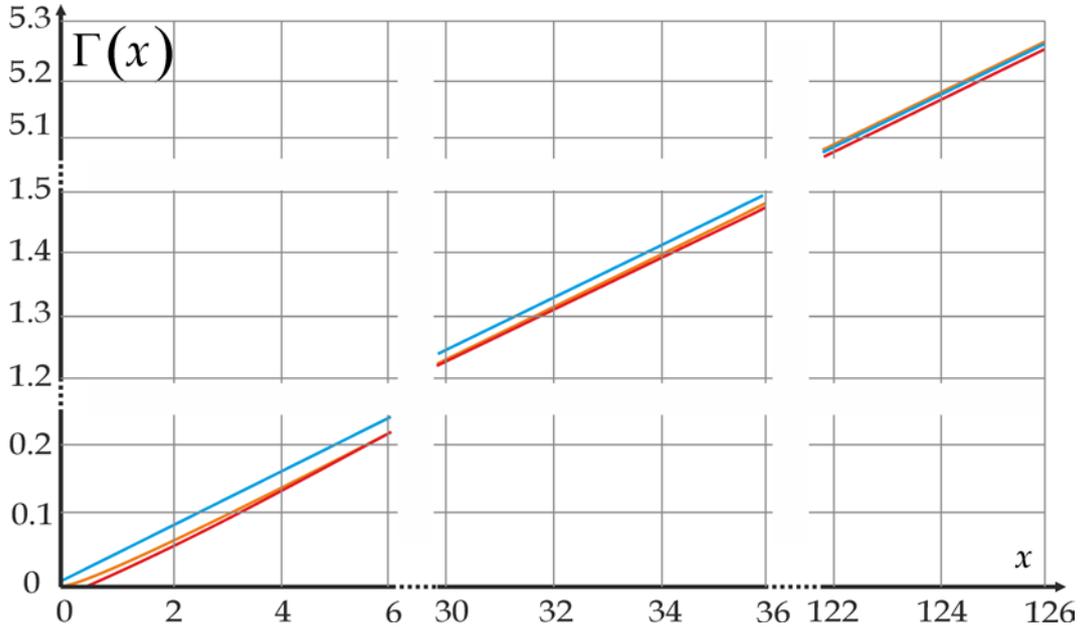

Fig.4. Fitting of $\Gamma_{MD}(t)$ (orange line) with the help of (13) (red line) and the simplified expression $\Gamma_{simp}(t) = C + 6D_s t$ (blue line) on several time intervals.

At the end of this Section we consider the fitting of $\Gamma_{MD}(t)$ with the help of (13) and the simplified expression $\Gamma_{simp}(t) = C + 6D_s t$. At small and intermediate times, the advantage of the formula (13) is evident (Fig. 4).

This circumstance can be considered as an additional evidence in the favor of the collective drift according to the described mechanism.

## 4. DISCUSSION OF THE RESULTS OBTAINED

The application of the Lagrange theory of thermal hydrodynamic fluctuation to the self-diffusion problem in liquid argon is considered in this work. It is shown that the functional structure of the mean square displacement for a molecule, predicted by the Lagrange theory, is in quite satisfactory agreement with the results of computer simulations. In particular, the following main results are obtained: 1) the existence of the predicted contribution to the MSD, proportional to $\sqrt{t}$, is established; 2) a new method for the determination of the Maxwell relaxation time is proposed. Values of $\tau_M$ on the coexistence curve are determined. With the good accuracy, $\tau_M$ proves to be close to constant and is in satisfactory agreement with the corresponding results of other works; 3) the relative values of the collective contribution to the self-diffusion coefficient are determined in two independent ways. It was shown that $D_c/D_s$ is about 20% near the triple point and it tends to 40 % at approaching the critical point (see Table II). At that, the estimates for $D_c/D_s$, obtained with the help of the MSD and the direct method, are consistent with each other. Taking into account this circumstance, we can estimate the ratio $D_c/D_s$ for arbitrary stable states of argon using the MSD analysis.

Besides, the jump-like changes of the self-diffusion coefficient $D_s$, $C$ and $t_a$ between $50K$ and $60K$ are observed. The last circumstance can be interpreted as a manifestation of the solid-liquid spinodal in argon. This result agrees satisfactory with the results of thermodynamic analysis in Ref. 30. Note, that all the main parameters that determine the behavior of the MSD are in good agreement with literature data.

Here it is necessary to make the following remark. From (1), (5) and (6) it follows that the long-time asymptote of $\phi_{\tilde{V}}(t)$ is determined by the equation:

$$\phi_{\tilde{V}}(t) = \frac{A}{\left(t + \Gamma(t)/6\nu\right)^{3/2}} \Rightarrow \frac{A}{t^{3/2}}\left(1 - \frac{1}{4}\frac{C}{\nu t} + ...\right).$$

Combining this result with (25) we find:

$$\phi_{\tilde{V}}(t) = \frac{A}{\left(t + \Gamma(t)/6\nu\right)^{3/2}} \Rightarrow \frac{A}{t^{3/2}}\left[1 - \frac{1}{4t}\left(\frac{C}{\nu} + 9\tau_M\right) + ...\right].$$

From here it follows that

$$\Delta\tau_M = -\frac{1}{9}\frac{C}{\nu}$$

However this change of the Maxwell relaxation time does not exceed (2-3) % for liquid argon.

We plan to apply our approach to other molecular liquids, in particular for water. In the last case the intermolecular potential is essentially more complicated. However for the description of slow self-diffusion process one can use the averaged intermolecular potential.

At the end of our paper we wish to thank Prof. L.A. Bulavin and Prof. G.G. Malenkov for the detailed discussion of the result obtained.

# APPENDIX: COMPUTER SIMULATIONS OF THE MSD FOR ARGON MOLECULES

For computer simulation of the MSD of a molecule we consider NVE ensemble consisting from $50^3 = 125000$ atoms of argon which occupy the cubic cell with the linear size $l_1 = 18.02\ nm$ at the temperature $T = 83.815\ K$ close to the triple point of argon. The periodic boundary conditions are used. To test the influence of the size of NVE ensemble we compare our results with those corresponding to the NVE ensemble of $10^6$ particles with $l_1 = 36\,nm$ (see Table AI). The Gromacs software environment was used for our purposes Refs. 31, 32.

TABLE AI. The characteristic spatial and time scales corresponding to the system

| N | $l_1, nm$ | $c, m/s$ | | $\tau_c, ps$ | |
|---|---|---|---|---|---|
| | | $T=84K$ | $T=150K$ | $T=84K$ | $T=150K$ |
| 125000 | 18.02 | 867 | 1041 | 20.8 | 17.3 |
| 1000000 | 36 | | | 46.1 | 38.4 |

Here, $c$ is the sound velocity and $\tau_c = l_1/c$ is the characteristic time during which we can ignore the interference effects due to periodic boundary conditions (PBC), $l_1$ is the size of the cubic cell. From the Table AI it follows that the time intervals used for determination of $D_c/D_s$ and MRT $\tau_M$ do not exceed 20 ps for the system of 125000 particles.

The additional verification is connected with the comparison of the self-diffusion coefficients calculated for two systems with different numbers of particles (Table AII).

TABLE AII. Values of the self-diffusion coefficient corresponding to two systems with different sizes: $D_s^{(1)}$ – for 125000 of particles, $D_s^{(2)}$ – for $10^6$ particles.

| $T, K$ | $D_s^{(1)} \cdot 10^{-5}, cm^2/s$ | $D_s^{(2)} \cdot 10^{-5}, cm^2/s$ |
|---|---|---|
| 95 | 2.67±0.127 | 2.82±0.04 |
| 110 | 4.43±0.186 | 4.62±0.078 |
| 120 | 5.91±0.141 | 5.84±0.162 |
| 130 | 7.73±0.207 | 7.81±0.112 |

As we can see differences between calculated values of the self-diffusion coefficient do not exceed statistical errors and that allows us to use the ensemble of $50^3$ particles for our purposes. The interaction between particles is described by the gromos54a7 force field Refs. 33, 34 that uses the potential:

$$U(r) = \frac{C_i^{(12)}}{r^{12}} - \frac{C_i^{(6)}}{r^6}, \tag{A1}$$

where $C_i^{(12)} = 0.9847 \cdot 10^{-5} \; kJ \cdot nm^{12} / mol$ and $C_i^{(6)} = 0.6265 \cdot 10^{-2} \; kJ \cdot nm^6 / mol$ Ref. 33

This potential is equivalent to

$$U(r) = 4\varepsilon \left[ \left(\frac{\sigma}{r}\right)^{12} - \left(\frac{\sigma}{r}\right)^6 \right], \tag{A2}$$

where $\varepsilon / k_B = 117.8 \; K$ and $\sigma = 3.405 \; \mathring{A}$.

The Verlet algorithm for the integration of the Newton equations with Verlet pair list was used. The intermolecular potential was shifted Ref. 35 on the distance $r_c = 7.5\sigma$, and the radius of pair list $r_l = 8\sigma$ was used. The integration step was never more than $0.1 \; fs$.

On the first simulation step the Maxwell velocity field for given temperature was generated. After relaxation process (more than 400 ps) oscillations of the potential energy become negligible and the soft thermostat procedure (the Nose-Hoover thermostat Refs. 36, 37) is used. The last stage of simulations includes 10 million time steps carried out without thermostat. Time dependencies of the MSD and VACF correspond to the averaging at all time steps.

The consistency of our simulations with those in recent work Ref. 39 is illustrated in Fig. A1.

The time dependencies of the MSD for an argon molecule are presented in Fig. A2.

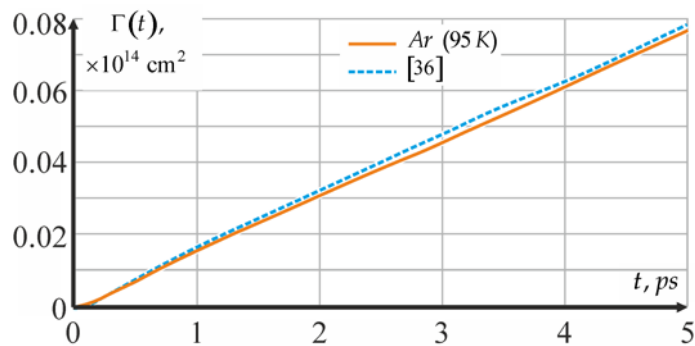

Fig. A1. The comparative behavior of the MSD obtained in our work and in Ref. 38 for $T = 95\,K$.

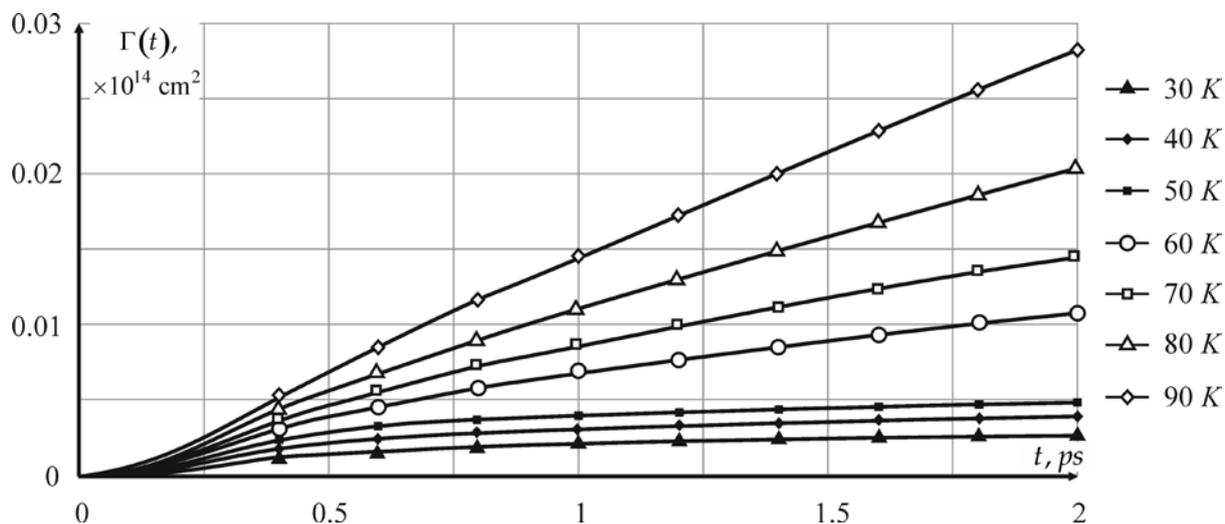

Fig. A2. The MSD for an argon molecule as functions of time and temperature. The dotted

lines are the linear extrapolations of the long-time asymptotes to small times.

Values of the density for temperatures higher that for the triple point are taken from NIST Tables Refs. 39-42 (one can also search data on the web "webbook.nist.gov/chemistry/fluid"). Its values for the supercooled region were obtained by the extrapolation according to the formulas describing the behavior of density for normal states (see Fig. A3).

The characteristic values of parameters determining the behavior of the MSD for a molecule are collected in the Table AIII. There the characteristic time $t_a$ is given. It separates the initial stage in time evolution of the MSD from the region of asymptotic behavior.

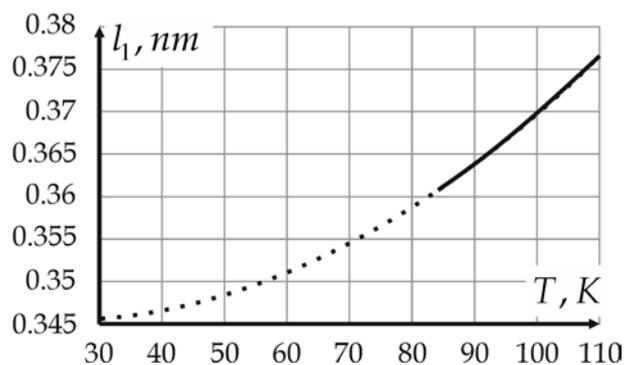

Fig. A3. The characteristic length $l_1 = n^{-1/3}$ as a function of temperature. The solid line corresponds to the NIST data Refs. 39-42, dotted line is obtained by the parabolic polynomial extrapolation.

TABLE AIII. The values of self-diffusion coefficient, constant $C$ and times $t_a$ at various temperatures

| $T, K$ | $t_a \cdot 10^{13}$, s | $C \cdot 10^{14}$, $cm^2$ | $D_s \cdot 10^{-5}$, $cm^2/s$ | $D_{exp} \cdot 10^{-5}$, $cm^2/s$ Refs. 43, 44 |
|---|---|---|---|---|
| 60 | 4.3 | 0.0032 | 0.4679 | – |
| 70 | 4.3 | 0.00235 | 0.8007 | – |
| 80 | 4.3 | 0.0012 | 1.4116 | – |
| 83.8 | 4.3 | 0.00097 | 1.6825 | 1.8 |
| 95 | 4.3 | 0.0002 | 2.67 | 2.73 |
| 110 | 4.3 | -0.0044 | 4.43 | 4.717 |
| 120 | 4.3 | -0.0124 | 5.91 | 6.179 |
| 130 | 4.6 | -0.0207 | 7.73 | 7.769 |

The approximation of the computer simulation results with the help of (17) for two temperatures $(T = 120\,K,\; T = 130\,K)$ is presented in Fig. A4.

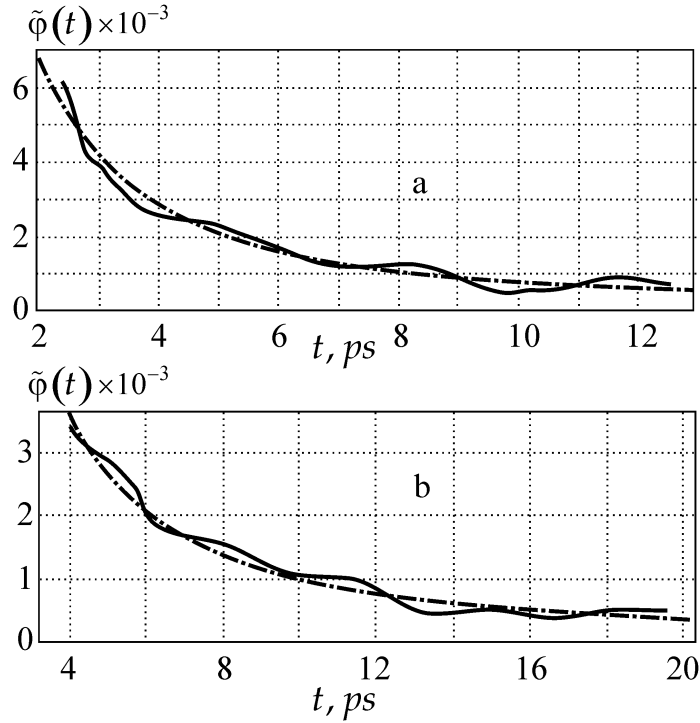

Fig. A4. Time dependencies of the VACF ($\tilde{\varphi}(t) = \varphi(t)/\varphi(0)$): solid lines present computer simulation data and dashed ones – interpolations with the help of (13) at $T = 120\,K$ (a) and $T = 130\,K$ (b)